\documentstyle[prd,eqsecnum,aps,floats,epsfig]{revtex}
\def\beq{\begin{eqnarray}}
\def\eeq{\end{eqnarray}}

\def\non{\nonumber}
\def\la{\langle}
\def\ra{\rangle}
\def\ep{\epsilon}

\def\Ova{O^q_{V-A}}
\def\Osp{O^q_{S-P}}
\def\Tva{T^q_{V-A}}
\def\Tsp{T^q_{S-P}}

\def\lsim{ {\ \lower-1.2pt\vbox{\hbox{\rlap{$<$}\lower5pt\vbox{\hbox{$\sim$}
}}}\ } }

\begin{document}

\draft
\vfill
\title{ Nonspectator Effects and $B$ Meson Lifetimes\\
}
\draft
\vfill
\author{
Kwei-Chou Yang \footnote{Email address: {\tt
kcyang@phys.sinica.edu.tw}\\
This work was done in collaboration with Hai-Yang Cheng}}
\address{Institute of Physics, Academia Sinica, Taipei, Taiwan 115, R.O.C.}

\maketitle
\date{June 1997}
\begin{abstract}
We review the $B$ meson lifetime problems and nonspectator
effects. The predictions of $B$ meson lifetime ratios depend on
four unknown hadronic parameters $B_1$, $B_2$, $\epsilon_1$ and
$\epsilon_2$, where $B_1$ and $B_2$ parametrize the matrix
elements of color singlet-singlet four-quark operators and
$\epsilon_1$ and $\epsilon_2$ the matrix elements of color
octet-octet operators. To understand contributions of the
nonspectator effects to the $B$ meson lifetime ratios, we derive
the renormalization-group improved QCD sum rules for these
parameters within the framework of heavy quark effective theory.
\end{abstract}

\vspace{0.7in}
\section{Introduction}

A QCD-based operator-product-expansion (OPE) formulation for
treatment of inclusive heavy hadron decays has been developed in
past years \cite{BIGI}. The optical theorem tell us that the
inclusive decay rates are related to the imaginary part of certain
forward scattering amplitudes along the physical cut. Since, based
on the hypothesis of quark-hadron duality, the final state effects
which are nonperturbative in nature, are eliminated after adding
up all of the states, the OPE approach thus can be employed for
such the smeared or averaged physical quantities. In order to test
the validity of (local) quark-hadron duality, it is very important
to have a reliable estimate of the heavy hadron lifetimes within
the OPE framework and compare them with experiment.

  In the heavy quark limit, all bottom hadrons have the same lifetimes
in the parton picture. With the advent of heavy quark effective
theory, which gives a systematic way in expansion of the initial
heavy hadron, and the OPE approach for the analysis of inclusive
weak decays, it is realized that the first nonperturbative
correction to bottom hadron lifetimes starts at order $1/m_b^2$.
However, the $1/m_b^2$ corrections are small and essentially
negligible in the lifetime ratios. The nonspectator effects such
as $W$-exchange and Pauli interference due to four-quark
interactions are of order $1/m_Q^3$, but their contributions can
be potentially significant due to a phase-space enhancement by a
factor of $16\pi^2$. As a result, the lifetime differences of
heavy hadrons come mainly from the above-mentioned nonspectator
effects.

\section{Difficulties of the OPE approach}

The world average lifetime ratios of bottom hadrons are
\cite{LEP}:
\begin{eqnarray}\label{taudata}
   {\tau(B^-)\over\tau(B^0_d)} &=& 1.07\pm 0.03 \,, \nonumber\\
   {\tau(B^0_s)\over\tau(B^0_d)} &=& 0.94\pm 0.04 \,, \nonumber\\
\frac{\tau(\Lambda_b)}{\tau(B^0_d)} &=& 0.79\pm 0.05 \,.
\end{eqnarray}
Since, to order $1/m_b^2$, the OPE results for all of the above
ratios are very close to unity [see Eq.~(\ref{taucrude}) below],
the conflict between theory and experiment for this lifetime ratio
is quite striking \cite{Uraltsev,NS,Cheng,BLLS}. One possible
reason for the discrepancy is that (local) quark-hadron duality
may not work in the study of nonleptonic inclusive decay widths.
Another possibility is that some hadronic matrix elements of
four-quark operators are probably larger than what naively
expected so that the nonspectator effects of order $16\pi^2/m_b^3$
may be large enough to explain the observed lifetime ratios.
Therefore, one cannot conclude that (local) duality truly fails
before a reliable calculation of the four-quark matrix elements is
obtained~\cite{NS}.

Conventionally, the hadronic matrix elements of four-quark
operators are evaluated using the factorization approximation for
mesons and the quark model for baryons. However, as we shall see,
nonfactorizable effects absent in the factorization hypothesis can
affect the $B$ meson lifetime ratios significantly. To have a
reliable estimate of the hadronic parameters $B_1$,
$B_2,~\epsilon_1$ and $\epsilon_2$ in the meson sector, to be
introduced below, we will apply the QCD sum rule to calculate
these unknown parameters.

\section{Theoretical review}
In this talk we will focus on the study of the four-quark matrix
elements of the $B$ meson. Before proceeding, let us briefly
review the theory. Applying the optical theorem, the inclusive
decay width of the hadron $H_b$ containing a $b$ quark can be
expressed as
\begin{equation}\label{imt}
   \Gamma(H_b\to X) = \frac{1}{m_{H_b}}\,\mbox{Im}\,
  i\!\int{\rm d}^4x\, \langle H_b|\,T\{\,
   {\cal L}_{\rm eff}(x),{\cal L}_{\rm eff}(0)\,\}
\,|H_b\rangle \,,
\end{equation}
where ${\cal L}_{\rm eff}$ is the relevant effective weak
Lagrangian that contributes to the particular final state $X$.
When the energy release in a $b$ quark decay is sufficiently
large, it is possible to express the nonlocal operator product in
Eq.~(\ref{imt}) as a series of local operators in powers  of $1/
m_b$ by using the OPE technique. In the OPE series, the only
locally gauge invariant operator with dimension four, $\bar b
i\!\! \not\!\! D b$, can be reduced to $m_b \bar bb$ by using the
equation of motion. Therefore, the first nonperturbative
correction to the inclusive $B$ hadron decay width starts at order
$1/m_b^2$. As a result, the inclusive decay width of a hadron
$H_b$ can be expressed as~\cite{BIGI}
\begin{eqnarray}\label{gener}
   \Gamma(H_b\to X) =&& \frac{G_F^2 m_b^5 |V_{\rm CKM}|^2}{192\pi^3}\,
\frac{1}{2m_{H_b}}
   \left\{  c_3^X\,\langle H_b|\bar b b|H_b\rangle
+ c_5^X\, \frac{\langle H_b|\bar b\,{1\over 2} g_s\sigma\cdot G b
|H_b\rangle} {m_b^2}\right. \nonumber\\
    && \left.+ \sum_n c_6^{X(n)}\,\frac{\langle H_b| O_6^{(n)}|H_b
    \rangle}{m_b^3} +  O(1/m_b^4) \right\} \,,
\end{eqnarray}
where $\sigma\cdot G=\sigma_{\mu\nu}G^{\mu\nu}$, $V_{\rm CKM}$
denotes some combination of the Cabibbo-Kobayashi-Maskawa
parameters and $c_i^X$ reflect short-distance dynamics and
phase-space corrections. The matrix elements in Eq.~(\ref{gener})
can be systematically expanded in powers of $1/m_b$ in heavy quark
effective theory (HQET), in which the $b$-quark field is
represented by a four-velocity-dependent field denoted by
$h^{(b)}_v(x)$. To first order in $1/m_b$, the $b$-quark field
$b(x)$ in QCD
 and the HQET-field $h^{(b)}_v(x)$ are related via
\begin{equation}\label{hqetb}
b(x) = e ^{-im_b v\cdot x} \left[ 1 + i\frac{\not\!\! D}{2m_b} \right]
h^{(b)}_v(x).
\end{equation}
Applying this relation, one can replace $b$ by the effective field
$h^{(b)}_v$ in Eq.~(\ref{gener}) to obtain
\begin{eqnarray}
  \frac{\langle H_b|\bar b b|H_b \rangle}{2m_{H_b}} &=& 1
    - \frac{K_{H_b}}{2m_b^2}+\frac{G_{H_b}}{2 m_b^2} + O(1/m_b^3) \,,
\nonumber\\
  \frac{\langle H_b|\bar b{1\over 2}g_s\sigma\cdot G b|H_b
  \rangle}{2m_{H_b}} &=& {G_{H_b}} + O(1/m_b) \,,
\end{eqnarray}
where
\begin{eqnarray}
   K_{H_b} \equiv -\frac{\langle H_b|\bar h^{(b)}_v\, (iD_\perp)^2
h^{(b)}_v |H_b \rangle}{2m_{H_b}}  \,,\ \ G_{H_b} \equiv
 \frac{\langle H_b|\bar h^{(b)}_v\,{1\over 2}g_s\sigma\cdot G
h^{(b)}_v |H_b \rangle}{2m_{H_b}} \,.
\end{eqnarray}
Note that here we adopt the convention
$D^\alpha=\partial^\alpha-ig_s A^\alpha$. The inclusive
nonleptonic and semileptonic decay rates of a bottom hadron to
order $1/m_b^2$ are given by \cite{BIGI} \beq \label{nl}
\Gamma_{\rm NL}(H_b) &=& {G_F^2m_b^5\over
192\pi^3}N_c\,|V_{cb}|^2\, {1\over 2m_{H_b}} \Bigg\{
\left(c_1^2+c_2^2+{2c_1c_2\over N_c}\right)\times \non \\ && \Big[
\big(\alpha I_0(x,0,0)+\beta I_0 (x,x,0)\big)\la H_b|\bar
bb|H_b\ra   \non \\ && -{1\over
m_b^2}\big(I_1(x,0,0)+I_1(x,x,0)\big) \la H_b|\bar bg_s\sigma
\cdot G b|H_b\ra \Big]   \non \\ && -{4\over m_b^2}\,{2c_1c_2\over
N_c}\,\big(I_2(x,0,0)+I_2(x,x,0)\big) \la H_b|\bar bg_s\sigma\cdot
G b|H_b\ra\Bigg\}, \eeq where $N_c$ is the number of colors, the
parameters $\alpha$ and $\beta$ denote QCD radiative corrections
to the processes $b\to c\bar ud$ and $b\to c\bar cs$, respectively
\cite{Bagan}, and \beq   \label{sl} \Gamma_{\rm SL}(H_b) &=&
{G_F^2m_b^5\over 192\pi^3}|V_{cb}|^2\,{ \eta(x,x_\ell,0)\over
2m_{H_b}} \nonumber \\ &\times& \Big[ I_0(x,0,0)\la H_b|\bar
bb|H_b\ra-{1\over m_b^2}\,I_1(x,0,0) \la H_b|\bar bg_s\sigma\cdot
G b|H_b\ra \Big]  \,, \eeq where $\eta(x,x_\ell,0)$ with
$x_\ell=(m_\ell/m_Q)^2$ is the QCD radiative correction to the
semileptonic decay rate and its general analytic expression is
given in \cite{Hokim}. In Eqs.~(\ref{nl}) and (\ref{sl}),
$I_{0,1,2}$ are phase-space factors (see e.g. \cite{Cheng} for
their explicit expressions): $I_i(x,0,0)$ for $b\to c\bar ud$
transition and $I_i(x,x,0)$ for $b\to c\bar cs$ transition. Note
that the CKM parameter $V_{ud}$ does not occur in $\Gamma_{\rm
NL}(H_b)$ and $\Gamma_{\rm SL}(H_b)$ when summing over the
Cabibbo-allowed and Cabibbo-suppressed contributions.

In Eq.~(\ref{nl}) $c_1$ and $c_2$ are the Wilson coefficients in the
effective Hamiltonian
\beq
{\cal H}^{\Delta B=1}_{\rm eff} &=& {G_F\over\sqrt{2}}\Big[ V_{cb}V_{uq}^*(c_1
(\mu)O_1^u(\mu)+c_2(\mu)O_2^u(\mu))   \nonumber \\
&+& V_{cb}V_{cq}^*(c_1(\mu)O_1^c(\mu)+c_2(\mu)
O_2^c(\mu))+\cdots \Big]+{\rm h.c.},
\eeq
where $q=d,s$, and
\beq \label{O12}
&& O_1^u= \bar c\gamma_\mu(1-\gamma_5)b\,\,\bar q\gamma^\mu(1-\gamma_5)u,
\qquad\quad
O_2^u =
\bar q\gamma_\mu(1-\gamma_5)b\,\,\bar c\gamma^\mu(1-\gamma_5)u \,.
\eeq
The scale and scheme dependence of the Wilson coefficients $c_{1,2}(\mu)$ are
canceled out by the corresponding dependence in the matrix element of the
four-quark operators $O_{1,2}$. That is, the four-quark operators
in the effective theory have to
be renormalized at the same scale $\mu$ and evaluated using the same
renormalization scheme as that for the Wilson coefficients.

Here we use the effective Wilson coefficients $c_i$ which are
scheme-independent~\cite{Cheng}. \beq \label{c12} c_1=1.149\,,
\qquad \quad c_2=-0.325\,. \eeq Using $m_b=4.85$ GeV, $m_c=1.45$
GeV, $|V_{cb}|=0.039$, $G_B=0.36\,{\rm GeV}^2$, $G_{\Lambda_b}=0$,
$K_B\approx K_{\Lambda_b}\approx 0.4\,{\rm GeV}^2$ together with
$\alpha=1.063$ and $\beta=1.32$ to the next-to-leading order
\cite{Bagan}, we find numerically
\begin{eqnarray}\label{taucrude}
   \frac{\tau(B^-)}{\tau(B_d)} = 1 + O(1/m_b^3)\,,\ \
   \frac{\tau(B_s)}{\tau(B_d)} = 1 + O(1/m_b^3)\,, \ \
   \frac{\tau(\Lambda_b)}{\tau(B_d)} =
   0.99 + O(1/m_b^3) \,.
\end{eqnarray}
It is evident that the $1/m_b^2$ corrections are too small to
explain the shorter lifetime of the $\Lambda_b$ relative to that
of the $B_d$. To the order of $1/m_b^3$, the nonspectator effects
due to Pauli interference and $W$-exchange parametrized in terms
of the hadronic parameters~\cite{NS}: $B_1$, $B_2$, $\epsilon_1$,
$\epsilon_2$, $\tilde B$, and $r$ (see below), may contribute
significantly to lifetime ratios due to a phase-space enhancement
by a factor of $16\pi^2$. The four-quark operators relevant to
inclusive nonleptonic $B$ decays are
\begin{eqnarray}\label{4qops}
   O_{V-A}^q &=& \bar b_L\gamma_\mu q_L\,\bar q_L\gamma^\mu b_L
    \,, \nonumber\\
   O_{S-P}^q &=& \bar b_R\,q_L\,\bar q_L\,b_R \,, \nonumber\\
   T_{V-A}^q &=& \bar b_L\gamma_\mu t^a q_L\,
\bar q_L\gamma^\mu  t^a b_L \,, \nonumber\\
   T_{S-P}^q &=& \bar b_R\,t^a q_L\,\bar q_L\, t^ab_R \,,
\end{eqnarray}
where $q_{R,L}={1\pm\gamma_5\over 2}q$. For the matrix elements of
these four-quark operators between $B$ hadron states, following,
\cite{NS} we adopt the definitions: \beq \label{parameters}
{1\over 2m_{ B_q}}\la \bar B_q|\Ova|\bar B_q\ra &&\equiv
{f^2_{B_q} m_{B_q} \over 8}B_1\,, \nonumber\\ {1\over 2m_{B_q}}\la
\bar B_q|\Osp|\bar B_q\ra &&\equiv {f^2_{B_q} m_{B_q}\over
8}B_2\,,\nonumber\\ {1\over 2m_{B_q}}\la \bar B_q|\Tva|\bar B_q\ra
&&\equiv {f^2_{B_q} m_{B_q}\over 8}\epsilon_1\,, \nonumber\\
{1\over 2m_{B_q}}\la \bar B_q|\Tsp|\bar B_q\ra &&\equiv {f^2_{B_q}
m_{B_q}\over 8}\epsilon_2\,,\nonumber\\ {1\over 2m_{\Lambda_b}}\la
\Lambda_b |\Ova|\Lambda_b \ra &&\equiv -{f^2_{B_q} m_{B_q}\over
48}r\,,\nonumber\\ {1\over 2m_{\Lambda_b}}\la \Lambda_b
|\Tva|\Lambda_b \ra &&\equiv -{1\over 2} (\tilde B+{1\over 3})
{1\over 2m_{\Lambda_b}}\la \Lambda_b |\Ova|\Lambda_b \ra \,. \eeq
Under the factorization approximation, $B_i=1$ and $\epsilon_i=0$,
and under the valence quark approximation $\tilde B=1$ \cite{NS}.

The destructive Pauli interference in inclusive nonleptonic $B^-$
decay and the $W$-exchange contributions to $B^0_d$ and $B^0_s$
are~\cite{NS} \footnote{The penguin-like nonspectator
contributions to $B_s$ are considered in \cite{Keum}, but they are
negligible compared to that from the current-current operators
$O_1$ and $O_2$ introduced in Eq.~(\ref{O12}).} \beq
\label{bnonspec} \Gamma^{\rm ann}(B^0_d) &=& -\Gamma_0
|V_{ud}|^2\, \eta_{\rm nspec}(1-x)^2\Bigg\{ (1+{1 \over
2}x)\Big[({1\over N_c}c_1^2+2c_1c_2+N_cc_2^2)B_1+2c_1^2\ep_1
\Big]\non \\ && -(1+2x)\Big[({1\over
N_c}c_1^2+2c_1c_2+N_cc_2^2)B_2 +2c_1^2\epsilon_2\Big]  \Bigg\}
\nonumber \\ &&-\Gamma_0 |V_{cd}|^2\, \eta_{\rm
nspec}\sqrt{1-4x}\Bigg\{ (1+{1 \over 2}x) \Big[({1\over
N_c}c_1^2+2c_1c_2+N_cc_2^2)B_1+2c_1^2\ep_1\Big]  \nonumber \\ &&
-(1+2x)\Big[({1\over N_c}c_1^2+2c_1c_2+N_cc_2^2)B_2
+2c_1^2\epsilon_2\Big]  \Bigg\}, \nonumber \\ \Gamma^{\rm
int}_-(B^-) &=& \Gamma_0\,\eta_{\rm nspec}(1-x)^2\left
[(c_1^2+c_2^2)(B_1+6\ep_1)+6c_1c_2B_1\right],\nonumber\\
\Gamma^{\rm ann}(B^0_s) &=& -\Gamma_0 |V_{cs}|^2\, \eta_{\rm
nspec}\sqrt{1-4x}\Bigg\{ (1+{1 \over 2}x) \Big[({1\over
N_c}c_1^2+2c_1c_2+N_cc_2^2)B_1+2c_1^2\ep_1\Big]  \nonumber \\ &&
-(1+2x)\Big[({1\over N_c}c_1^2+2c_1c_2+N_cc_2^2)B_2
+2c_1^2\epsilon_2\Big]  \Bigg\}  \nonumber \\ && -\Gamma_0
|V_{us}|^2\, \eta_{\rm nspec}(1-x)^2\Bigg\{ (1+{1 \over
2}x)\Big[({1\over N_c}
c_1^2+2c_1c_2+N_cc_2^2)B_1+2c_1^2\ep_1\Big]\nonumber\\ &&
-(1+2x)\Big[({1\over N_c}c_1^2+2c_1c_2+N_cc_2^2)B_2
+2c_1^2\epsilon_2\Big]  \Bigg\}    \,, \eeq with \beq
\Gamma_0={G_F^2m_b^5\over 192\pi^3}|V_{cb}|^2,~~~\eta_{\rm
nspec}=16 \pi^2{f_{B_q}^2m_{B_q}\over m_b^3}\,, \eeq where
$f_{B_q}$ is the $B_q$ meson decay constant defined by
\begin{equation}
\langle 0|\bar q\gamma_\mu \gamma_5 b|\bar B_q(p)\rangle =if_{B_q} p_\mu \,.
\end{equation}
Likewise, the nonspectator effects in inclusive nonleptonic decays
of the $\Lambda_b$ baryon are given by \cite{NS} \beq
\label{lnonspec} \Gamma^{\rm ann}(\Lambda_b) &=& {1\over
2}\Gamma_0\,\eta_{\rm nspec} \,r(1-x)^2\Big (\tilde
B(c_1^2+c_2^2)-2c_1c_2\Big),   \\ \Gamma^{\rm int}_-(\Lambda_b)
&=& -{1\over 4}\Gamma_0 \, \eta_{\rm nspec}\,r
\left[|V_{ud}|^2(1-x)^2(1+x)+\left|{V_{cd}}\right|^2\sqrt{1-4x}\,\right]
\Big(\tilde Bc_1^2-2c_1c_2-N_cc_2^2\Big)\,.   \nonumber \eeq
Using
the values of $c_i$ in Eqs.~(\ref{c12}), we obtain
  \beq \label{numnspec} \Gamma^{\rm ann}(B_d) &=&
\Gamma_0\, \eta_{\rm nspec} (-0.0087 B_1+0.0098 B_2 -2.28
\epsilon_1 +2.58\epsilon_2)\,, \nonumber\\ \Gamma_-^{\rm int}(B^-)
&=& \Gamma_0 \, \eta_{\rm nspec}(-0.68B_1 +7.10 \epsilon_1)\,,
\nonumber\\ \Gamma^{\rm ann}(B_s) &=& \Gamma_0 \, \eta_{\rm
nspec}(-0.0085 B_1 +0.0096 B_2 -2.22 \epsilon_1 +2.50 \epsilon_2)
\,, \nonumber\\ \Gamma^{\rm ann}(\Lambda_b) &=& \Gamma_0 \,
\eta_{\rm nspec}r (0.59 \tilde B +0.31)\,, \nonumber\\ \Gamma^{\rm
int}(\Lambda_b) &=& \Gamma_0 \, \eta_{\rm nspec}r (-0.30 \tilde B
-0.097) \,. \eeq Therefore, to the order of $1/m_b^3$, the
$B$-hadron lifetime ratios are given by
\begin{eqnarray}\label{ratios}
\frac{\tau(B^-)}{\tau(B^0_d)} &=& 1 +
   \Bigl( {f_B\over 185~{\rm MeV}} \Big)^2 \Big( 0.043 B_1 + 0.0006 B_2
    - 0.61 \epsilon_1 + 0.17 \epsilon_2 \Big) \,, \nonumber \\
\frac{\tau (B^0_s)}{\tau(B^0_d)} &=& 1+ \Big(\frac{f_B}{185~{\rm
MeV}}\Bigr)^2 (-1.7\times 10^{-5}\,B_1+1.9\times 10^{-5}\, B_2
\,-0.0044 \epsilon_1\, +0.0050\, \epsilon_2)  \,, \nonumber\\
   \frac{\tau(\Lambda_b)}{\tau(B^0_d)} &=& 0.99
   +  \Bigl( {f_B\over 185~{\rm MeV}} \Big)^2 \Big[ -0.0006 B_1
    + 0.0006 B_2   \nonumber \\
&&   - 0.15 \epsilon_1 + 0.17 \epsilon_2
    - (0.014 + 0.019 \widetilde B) r \Big] \,.
\end{eqnarray}
We see that the coefficients of the color singlet--singlet
operators are one to two orders of magnitude smaller than those of
the color octet--octet operators. This implies that even a small
deviation from the factorization approximation $\epsilon_i=0$ can
have a sizable impact on the lifetime ratios. It was argued in
\cite{NS} that the unknown nonfactorizable contributions render it
impossible to make reliable estimates on the magnitude of the
lifetime ratios and even the sign of corrections. That is, the
theoretical prediction for $\tau(B^-)/\tau(B_d)$ is not
necessarily larger than unity. In the next section we will apply
the QCD sum rule method to estimate the aforementioned hadronic
parameters, especially $\epsilon_i$.

\section{The QCD sum rule calculation}
    In HQET where the $b$ quark is
treated as a static quark,
we can use the renormalization group equation to express
them in terms of the operators renormalized at a scale
 $ \Lambda_{\rm QCD}\ll \mu\ll m_b$. Their
renormalization-group evolution is determined by the ``hybrid"
anomalous dimensions~\cite{SV1} in HQET. The operators $O_{V-A}^q$
and $T_{V-A}^q$, and similarly $O_{S-P}^q$ and $T_{S-P}^q$, mix
under renormalization. In the leading logarithmic approximation,
the renormalization-group equation of the operator pair $(O,T)$
reads
\begin{equation}\label{rge}
\frac{d}{dt}
\left(
   \begin{array} {cc} \phantom{ \bigg[ } O \\
   T\end{array} \right)
    = {3\alpha_s\over 2\pi}\,\left(
   \begin{array} {cc} \phantom{ \bigg[ } C_F & -1 \\
    \displaystyle -{C_F\over 2 N_c}~ & \displaystyle ~{1\over 2 N_c}
   \end{array} \right)
\left(
   \begin{array} {cc} \phantom{ \bigg[ } O \\
   T\end{array} \right)
 \,,
\end{equation}
where $t = \frac{1}{2} \ln(Q^2/\mu^2)$,
$C_F = (N_c^2-1)/2 N_c$, and effects of penguin operators induced
from evolution have been neglected.

The solution to the evolution equation
Eq.~(\ref{rge}) has the form
\begin{equation}\label{diag}
\left(
   \begin{array} {cc} O \\
   T\end{array} \right)_{Q}
    = \left(
   \begin{array} {cc} \phantom{ \bigg[ } \frac{8}{9}~ & \frac{2}{3} \\
    -\frac{4}{27}~ &
\frac{8}{9}
   \end{array} \right)
\left(
   \begin{array} {cc} \phantom{ \bigg[ } L_{Q}^{9/(2\beta_0)}~ & 0 \\
    0~ & 1
   \end{array} \right)
 {\bf D_\mu}\,,
\end{equation}
where
\begin{equation}\label{d}
{\bf D_\mu}=
\left(
   \begin{array} {cc} D_1 \\
   D_2\end{array} \right)_\mu
    = \left(
   \begin{array} {cc} \phantom{ \bigg[ } O-\frac{3}{4}T \\
   \frac{1}{6}O+T\end{array} \right)_\mu
 \,,
\end{equation}
$L_Q = {\alpha_s(\mu)/\alpha_s(Q)}$ and
$\beta_0=\frac{11}{3}\,N_c-\frac{2}{3}\,n_f$ is the leading-order
expression of the $\beta$-function with $n_f$ being the number of
light quark flavors. The subscript $\mu$ in Eq.~(\ref{d}) and in
what follows denotes the renormalization point of the operators.
Given the evolution equation (\ref{diag}) for the four-quark
operators, we see that the hadronic parameters $B_i$ and
$\epsilon_i$ normalized at the scale $m_b$ are related to that at
$\mu=1$~GeV by
\begin{eqnarray}\label{lowpara}
   B_i(m_b) &\simeq& 1.54 B_i(\mu)
    - 0.41\epsilon_i(\mu) \,, \nonumber\\
   \epsilon_i(m_b) &\simeq& - 0.090 B_i(\mu)+1.07\epsilon_i(\mu) \,,
\end{eqnarray}
with $\mu=1$ GeV, where uses have been made of
$\alpha_s(m_{\rm Z})=0.118$,
$\Lambda^{(4)}_{\over{\rm MS}}=333~{\rm MeV}$, $m_b=4.85~{\rm GeV}$,
$m_c=1.45$ GeV.
The above results (\ref{lowpara}) indicate that renormalization effects
are quite significant.

It is easily seen from Eqs.~(\ref{diag}) and (\ref{d}) that the
normalized operator $D_1$ (or $D_2$) is simply multiplied by
$L_{Q}^{9/(2\beta_0)}$ (or 1) when it evolves from a
renormalization point $\mu$ to another point $Q$. In what
follows\footnote{In the sum rule calculation, the factorization
scale $\mu$ cannot be chosen too small, otherwise the strong
coupling constant $\alpha_s$ would be so large that Wilson
coefficients cannot be perturbatively calculated.}, we will apply
this  property to derive the renormalization-group improved QCD
sum rules for $D_j$ at the typical scale $\mu=1$~GeV. We define
the new four-quark matrix elements as follows \beq {1\over 2m_{
B_q}}\la \bar B_q|D_j^{(i)}(\mu)|\bar B_q\ra \equiv {f^2_{B_q}
m_{B_q}\over 8}\, d_j^{(i)}(\mu), \eeq where the superscript $(i)$
denotes $(V-A)$ four-quark operators for $i=1$ and $(S-P)$
operators for $i=2$, and $d_j^{(i)}$ satisfy
\begin{equation}
\left(
   \begin{array} {cc} d_1^{(i)} \\
   d_2^{(i)}\end{array} \right)_\mu
    = \left(
   \begin{array} {cc} \phantom{ \bigg[ } B_i-\frac{3}{4}\epsilon_i \\
   \frac{1}{6}B_i+\epsilon_i\end{array} \right)_\mu
 \,.
\end{equation}

Since the terms linear in four-quark matrix elements are already
of order $1/m_b^3$, we  only need the relation between the full
QCD field $b(x)$ and the HQET field $h^{(b)}_v(x)$ to the zeroth
order in $1/m_b$: $b(x) = e ^{-im_b v\cdot x}\, \{h^{(b)}_v(x) +
{\cal O}(1/m_b)\}$. In the following, within the framework of
HQET, we apply the method of QCD sum rules to obtain the value of
the matrix elements of four-quark operators. We consider the
three-point correlation function \beq \label{corr}
\Pi^{D_j^{v(i)}}_{\alpha,\beta}(\omega,\omega')=i^2\int dx\, dy\,
e^{i\omega v\cdot x-i\omega' v\cdot y} \la 0|T\{[\bar
q(x)\Gamma_\alpha h^{(b)}_v(x)]\, D_j^{v(i)}(0)\, [\bar
q(y)\Gamma_\beta h^{(b)}_v(y)]^\dagger\}|0\ra \,, \eeq where the
operator $D_j^{(i)}$ is defined in Eq.~(\ref{d}) but with $b\to
h_v^{(b)}$ and $\Gamma_\alpha$ is chosen to be $v_\alpha\gamma_5$
(some further discussions can be found in \cite{HY}).

The correlation function can be written in the double dispersion
relation form \beq
\Pi^{D_j^{v(i)}}_{\alpha,\beta}(\omega,\omega')=\int\int {ds\over
s-\omega}\, {ds'\over s'-\omega'}\, \rho^{D_j^{v(i)}} \,. \eeq

The results of the QCD sum rules are obtained in the following
way. On the phenomenological side, which is the sum of the
relevant hadron states, this correlation function can be written
as \beq \Pi^{PS}_{D_j^{v(i)}}(\omega,\omega')=
\frac{F^2(m_b)F^2(\mu)d_j^{(i)}} {16(\bar \Lambda -\omega)(\bar
\Lambda -\omega')}+\cdots \,, \eeq where $\bar\Lambda$ is the
binding energy of the heavy meson in the heavy quark limit and
ellipses denote resonance contributions. The
heavy-flavor-independent decay constant $F$ defined in the heavy
quark limit is given by \beq \langle 0|\bar q\gamma^\mu\gamma_5
h^{(b)}_v|\bar B(v)\rangle =iF(\mu) v^\mu\,. \eeq The decay
constant $F(\mu)$ depends on the scale $\mu$ at which the
effective current operator is renormalized and it is related to
the scale-independent decay constant $f_B$ of the $B$ meson by
\beq F(m_b)=f_B\,\sqrt{m_B}. \eeq

On the theoretical side, the correlation function can be
alternatively calculated in terms of quarks and gluons using the
standard OPE technique.  Then we equate the results on the
phenomenological side with that on the theoretical side. However,
since we are only interested in the properties of the ground state
at hand, e.g., the $B$ meson, we shall assume that contributions
from excited states (on the phenomenological side) are
approximated by the spectral density on the theoretical side of
the sum rule, which starts from some thresholds (say,
$\omega_{i,j}$ in this study). To further improve the final result
under consideration, we apply the Borel transform to both external
variables $\omega$ and $\omega'$. After the Borel
transform~\cite{yang1}, \beq {\bf
B}[\Pi^{D_j^{v(i)}}_{\alpha,\beta}(\omega,\omega')]=
\lim_{{\scriptstyle m\to \infty \atop\scriptstyle -\omega'\to
\infty} \atop\scriptstyle {-\omega'\over mt'}\ {\rm fixed}}
\lim_{{\scriptstyle n\to \infty \atop\scriptstyle -\omega\to
\infty} \atop\scriptstyle {-\omega\over nt}\ {\rm fixed}} {1\over
n!m!}(-\omega')^{m+1} [{d\over d\omega'}]^m (-\omega)^{n+1}
[{d\over d\omega}]^n
\Pi^{D_j^{v(i)}}_{\alpha,\beta}(\omega,\omega')\,, \eeq the sum
rule gives \beq &&\frac{F^2({m_b}) F^2(\mu)}{16} e^{-
\bar\Lambda/t_1} e^{-\bar\Lambda/t_2} d_j^{(i)}
=\int_0^{\omega_{i,j}} ds\int_0^{\omega_{i,j}} ds' e^{-(s/t_1 +
s'/t_2)}\rho^{\rm QCD}\,, \eeq where $\omega_{i,j}$ is the
threshold of the excited states and $\rho^{\rm QCD}$ is the
spectral density on the theoretical side of the sum rule. Because
the Borel windows are symmetric in variables $t_1$ and $t_2$, it
is natural to choose $t_1=t_2$. However, unlike the case of the
normalization of the Isgur-Wise function at zero recoil, where the
Borel mass is approximately twice as large as that in the
corresponding two-point sum rule~\cite{Neubert2}, in the present
case of the three-point sum rule at hand, we find that the working
Borel windows can be chosen as the same as that in the two-point
sum rule since in our analysis the output results depend weakly on
the Borel mass. Therefore, we choose $t_1=t_2=t$.
 By the renormalization group technique, the logarithmic dependence
$\alpha_s\ln (2t/\mu)$ can be summed over to produce a factor like
$[\alpha_s(\mu)/\alpha_s(2t)]^\gamma$. After some manipulation we
obtain the sum rule results:
\begin{eqnarray}\label{rule1}
&&\frac{F^2({m_b}) F^2(\mu)}{16} e^{-2 \bar\Lambda/t} \left(
   \begin{array} {cc} \phantom{ \bigg[ } d_1^{v(i)} \\
   d_2^{v(i)}\end{array} \right)_\mu  \nonumber\\
=&& \Biggl( {\alpha_s(2t)\over \alpha_s(\mu)}
\Biggr)^{4\over\beta_0} \Biggl(
{1-2\delta{\alpha_s(2t)\over\pi}\over 1-2\delta{\alpha_s(\mu)
\over\pi}}\Biggr)
 \left(
   \begin{array} {cc} L_{t}^{-9/(2\beta_0)}~ & 0 \\
    0~ & 1
   \end{array} \right)
 \left(
   \begin{array} {cc} \phantom{ \bigg[ } {\rm OPE}_{B_{i,1}}
   -\frac{3}{4}\,{\rm OPE}_{\epsilon_{i,1}} \\
    \frac{1}{6}\, {\rm OPE}_{B_{i,2}}+{\rm OPE}_{\epsilon_{i,2}}
   \end{array}
   \right)_t \,,
\end{eqnarray}
where
\begin{eqnarray}\label{rule2}
&&{\rm OPE}_{B_{i,j}}\simeq \frac{1}{4}({\rm OPE})^2_{2pt;i,j} \,,
\nonumber\\ && {\rm OPE}_{\epsilon_{1,j}} \simeq
-\frac{1}{16}\Biggl[- \frac{ \langle \bar q g_s\sigma\cdot G
q\rangle}{8\pi^2} t (1-e^{-\omega_{1,j} /t})+\frac {\langle
\alpha_s G^2\rangle} {16\pi^3} t^2 (1-e^{-\omega_{1,j} /t})^2
\Biggr] \,, \nonumber\\ && {\rm OPE}_{\epsilon_{2,j}} \simeq {\cal
O}(\alpha_s)\,,
\end{eqnarray}
with
\begin{eqnarray} \label{2pt}
({\rm OPE})_{2pt;i,j} =&& \frac{1}{2}  \biggl\{
\int_0^{\omega_{i,j}} ds\ s^2 e^{-s/t}{3\over\pi^2}
\biggl[1+{\alpha_s\over\pi} \Bigl({17\over 3}+{4\pi^2\over 9}-2\ln
{s\over t}\Bigr) \biggr]\nonumber\\
&&-\Bigl(1+{2\alpha_s\over\pi}\Bigr)\langle\bar qq\rangle
+{\langle\bar qg_s \sigma\cdot Gq\rangle\over 16 t^2} \biggr\} \,.
\end{eqnarray}
For reason of consistency, in the following numerical analysis we
will neglect the finite part of radiative one loop corrections in
OPE$_{B_{i,j}}$ and OPE$_{\epsilon_{i,j}}$ (and in Eq. (\ref{F})).
The parameter $\delta$ in (\ref{rule1}) is some combination of the
$\beta$ functions and anomalous dimensions (see Eq.~(4.2) of
\cite{BB}) and is numerically equal to $-0.23$. The relevant
parameters normalized at the scale $t$ are related to those at
$\mu$ by~\cite{BB,yang1} \beq\label{RGevo} &&F(2t)=F(\mu)\Bigl(
{\alpha_s(2t)\over \alpha_s(\mu)} \Bigr)^{-2/\beta_0}
{1-\delta{\alpha_s(\mu)\over\pi} \over
1-\delta{\alpha_s(2t)\over\pi}}\,, \nonumber\\ &&\langle \bar
qq\rangle_{2t} =\langle \bar qq\rangle_\mu \cdot \Bigl(
{\alpha_s(2t)\over \alpha_s (\mu)}
\Bigr)^{-4/\beta_0}\,,\nonumber\\ &&\langle g_s\bar q\sigma\cdot
Gq\rangle_{2t}=\langle g_s\bar q\sigma\cdot Gq\rangle_\mu \cdot
\Bigl( {\alpha_s(2t)\over \alpha_s(\mu)} \Bigr)^{2/(3\beta_0)}
\,,\nonumber\\ &&\langle \alpha_s G^2 \rangle_{2t}= \langle
\alpha_s G^2 \rangle_\mu\,, \eeq where $\langle\cdots \rangle$
stands for $\langle 0| \cdots |0\rangle$ and~\cite{yang1} \beq
&&\langle \bar qq\rangle_{\mu=1~{\rm GeV}}=-(240~{\rm MeV})^3\,,
\nonumber\\ &&\langle \alpha_s G^2 \rangle_{\mu=1~{\rm GeV}}
=0.0377~{\rm GeV^4} \,,\nonumber\\ &&\langle \bar
qg_s\sigma_{\mu\nu} G^{\mu\nu} q\rangle_{\mu=1~{\rm GeV}}=
(0.8~{\rm GeV^2})\times \langle \bar qq\rangle_{\mu=1~{\rm GeV}}
\,. \eeq

Some remarks are in order. First, in Eqs.~(\ref{rule1}) and
(\ref{rule2}). OPE$_{B_i}$ is obtained by substituting
$D_j^{v(i)}$ by $O^v$  and it can be approximately factorized as
the product of (OPE)$_{2pt;i,j}$ with itself, which is the same as
the theoretical part in the two-point $F(\mu)$ sum
rule~\cite{Neubert2,BB}.  In the series of (OPE)$_{2pt;i,j}$, we
have neglected the contribution proportional to $\langle \bar
qq\rangle^2$. (More precisely, it is equal to $\alpha_s \langle
\bar qq\rangle^2 \pi/324$; see Ref.~\cite{Neubert2}.)
Nevertheless, the result of (OPE)$_{B_i}$ in Eq.~(\ref{rule2}) is
reliable up to dimension six, as the contributions from the
$\langle \bar qq\rangle^2$ terms in (OPE)$_{2pt;i,j}$ are much
smaller than the term $(1+\alpha_s/\pi)^2 \langle \bar qq
\rangle^2/16$ that we have kept [see Eq.~(\ref{2pt})]. Second, in
(OPE)$_{B_i}$ the contribution involving the gluon condensate is
proportional to the light quark mass and hence can be neglected.
Third, OPE$_{\epsilon_i}$ is the theoretical side of the sum rule,
and it is obtained by substituting $D_j^{v(i)}$ by $T^v$. Here we
have neglected the dimension-6 four-quark condensate of the type
$\langle \bar q\Gamma\lambda^a q\ \bar q\Gamma\lambda^a q
\rangle$. Its contribution is much less than that from
dimension-five or dimension-four condensates and hence unimportant
(see~\cite{Chern} for similar discussions). It should be
emphasized that nonfactorizable contributions to the parameters
$B_i$ arise mainly from the $O^v-T^v$ operator mixing.

In the following, we compare our analysis with the similar QCD sum
rule studies in \cite{Chern} and \cite{BLLS}. First, Chernyak
\cite{Chern} used the chiral interpolating current for the $B$
meson, so that all light quark fields in his correlators are
purely left-handed. As a result, there are no quark-gluon mixed
condensates as these require the presence of both left- and
right-handed light quark fields. Instead, the gluon condensate
contribution enters into the $\epsilon_1$ sum rule with an
additional factor of 4 in comparison with ours; thus their
$OPE_{\epsilon_1}$ is in rough agreement with ours. Second, our
results for OPE$_{\epsilon_i}$ are very different from that
obtained by Baek {\it et al}.~\cite{BLLS}. The reason is that
their results are mixed with the $1^+$ to $1^+$ transitions. Also
a subtraction of the contribution from excited states is not
carried out in \cite{BLLS} for the three-point correlation
function, though it is justified to do so for two-point
correlation functions. Indeed, in the following analysis, one will
find that after subtracting the contribution from excited states,
the contributions of OPE$_{\epsilon_i}$ are largely suppressed.
Furthermore, as in the study of the $B$ meson decay
constant~\cite{Neubert2}, we find that the renormalization-group
effects are very important in the sum rule analysis.  Moreover,
$\epsilon_i$ at $\mu=m_b$ are largely enhanced by
renormalization-group effects.

The value of $F$ in Eq.~(\ref{rule1}) can be substituted by
\begin{eqnarray}\label{F}
F^2(\mu)e^{-\bar\Lambda/t}=&& \biggl[{\alpha_s(2t)\over
\alpha_s(\mu)}\biggr]^{4\over\beta}
\biggl[{1-2\delta{\alpha_s(2t)\over\pi}\over
1-2\delta{\alpha_s(\mu)\over\pi}} \biggr]  \biggl\{
\int_0^{\omega_0} ds\ s^2 e^{-s/ t}{3\over\pi^2}
\biggl[1+{\alpha_s(2t)\over\pi} \Bigl({17\over 3}+{4\pi^2\over
9}-2\ln {s\over t}\Bigr) \biggr]\nonumber\\
&&-\Bigl(1+{2\alpha_s(2t)\over\pi}\Bigr)\langle\bar qq\rangle_{2t}
+{\langle\bar qg_s \sigma\cdot Gq\rangle_{2t}\over 16 t^2}
\biggr\} \,,
\end{eqnarray}
which is from the two-point sum rule approach~\cite{BB}. Next, to
determine the thresholds $\omega_{i,j}$ we employ the $B$ meson
decay constant $f_B=(185\pm 25\pm 17)~{\rm MeV}$ obtained from a
recent lattice-QCD calculation~\cite{lattice} and the
relation~\cite{Ball} \beq f_B ={F(m_b)\over \sqrt {m_B}}\Bigl(
1-{2\over 3} {\alpha_s(m_b)\over \pi }\Bigr) \Bigl(1-{(0.8\sim 1.1
)~{\rm GeV}\over m_b} \Bigr)\,, \eeq that takes into account QCD
and $1/m_b$ corrections. Using the relation between $F(m_b)$ and
$F(\mu)$ given by Eq.~(\ref{RGevo}) and $m_b=(4.85\pm 0.25)$~GeV,
we obtain \beq \label{Fresult} F(\mu=1~{\rm GeV}) \cong (0.34\sim
0.52)~{\rm GeV^{3/2}} \,. \eeq Since the $\bar\Lambda$ parameter
in Eq.~(\ref{F}) can be replaced by the $\bar\Lambda$ sum rule
obtained by applying the differential operator $t^2 \partial
\ln/\partial t$ to both sides of Eq.~(\ref{F}), the $F(\mu)$ sum
rule can be rewritten as \beq  \label{newF} F^2(\mu)={\rm (right\
hand\ side\ of\ Eq.~(\ref{F}))} \times {\rm exp} [t\,
{\partial\over \partial t} {\rm ln}{\rm (right\ hand\ side\ of\
Eq.~(\ref{F})) }]\,, \eeq which is $\bar\Lambda$-free. Then using
the result (\ref{Fresult}) as  input, the threshold $\omega_0$ in
the $F(\mu)$ sum rule, Eq.~(\ref{newF}), is determined. The result
for $\omega_0$ is  $1.25-1.65~{\rm GeV}$. A larger
$F(\mu=1~\rm{GeV})$ corresponds to a larger $\omega_0$. The
working Borel window lies in the region $0.6~{\rm GeV} <t < 1~{\rm
GeV}$, which turns out to be a reasonable choice. Substituting the
value of $\omega_0$ back into the $\bar\Lambda$ sum rule, we
obtain
 $\bar \Lambda=0.48-0.76~{\rm GeV}$ in the Borel window
$0.6~{\rm GeV} <t < 1~{\rm GeV}$. This result is consistent with
the choice $m_b=(4.85\pm 0.25)$~GeV, recalling that in the heavy
quark limit, $\bar\Lambda=m_B-m_b$. To extract the $d_j^{v(i)}$
sum rules, one can take the ratio of Eq.~(\ref{F}) and
Eq.~(\ref{rule1}) to eliminate the contribution of $F^2/ {\rm
exp}(\bar \Lambda /t)$. This means one has chosen the same $\bar
\Lambda$ both in Eq.~(\ref{F}) and Eq.~(\ref{rule1}). Since
quark-hadron duality is the basic assumption in the QCD sum rule
approach, we expect that the same result of $\bar \Lambda$ also
can be obtained using the $\bar\Lambda$ sum rules derived from
Eq.~(\ref{rule1}) (see \cite{yang1} for a further discussion).
This property can help us to determine consistently the threshold
in 3-point sum rule, Eq.~(\ref{rule1}). Therefore, we can apply
the differential operator $t^2 \partial \ln/\partial t$ to both
sides of Eq.~(\ref{rule1}), the $d^{v(i)}$ sum rule, to obtain new
$\bar \Lambda$ sum rules. The requirement of producing a
reasonable value for $\bar \Lambda$, say $0.48-0.76~{\rm GeV}$,
provides severe constraints on the choices of $\omega_{i,j}$. With
a careful study, we find that the best choice in our analysis is
\beq \label{omega3pt} \omega_{i,1}=-0.02~{\rm GeV} +\omega_0\,,
\quad \omega_{1,2}=-0.5~{\rm GeV}+\omega_0 \,, \quad
\omega_{2,2}=-0.22~{\rm GeV}+\omega_0 \,. \eeq Applying the above
relations with $\omega_0=(1.25 \sim 1.65)~{\rm GeV}$ and
substituting $F(\mu)$ in Eq.~(\ref{rule1}) by (\ref{F}), we study
numerically the $d_j^{v(i)}$ sum rules. In Fig.~1, we plot $B_i^v$
and $\epsilon_i^v$ as a function $t$, where $B_i^v=8d_1^{v(i)}/9 +
2d_2^{v(i)}/3,\ $ and $\epsilon_i^v=-4d_1^{v(i)}/27+
8d_2^{v(i)}/9$. The dashed and solid curves stand for $B_i^{v}$
and $\epsilon_i^{v}$, respectively, where we have used
$\omega_0=1.4~ {\rm GeV}$ (the corresponding decay constant is
$f_B=175\sim 195~ {\rm MeV}$ or $F(\mu)=0.405\pm 0.005~ {\rm
GeV}^{3/2}$). The final results for the hadronic parameters $B_i$
and $\epsilon_i$ are (see Fig.~2) \beq
 B_1^{v}(\mu=1~{\rm GeV})=0.60\pm 0.02, \qquad &&
B_2^{v}(\mu=1~{\rm GeV})=0.61\pm 0.01, \nonumber \\
 \epsilon_1^{v}(\mu=1~{\rm GeV})=-0.08\pm 0.01,   \qquad  &&
\epsilon_2^{v}(\mu=1~{\rm GeV})=-0.024\pm 0.006. \eeq The
numerical errors come mainly from the uncertainty of
$\omega_0=1.25\sim 1.65$~GeV. Some intrinsic errors of the sum
rule approach, say quark-hadron duality or $\alpha_s$ corrections,
will not be considered here.

\begin{figure}[ht]
\vspace{1cm}
    \leftline{\epsfig{figure=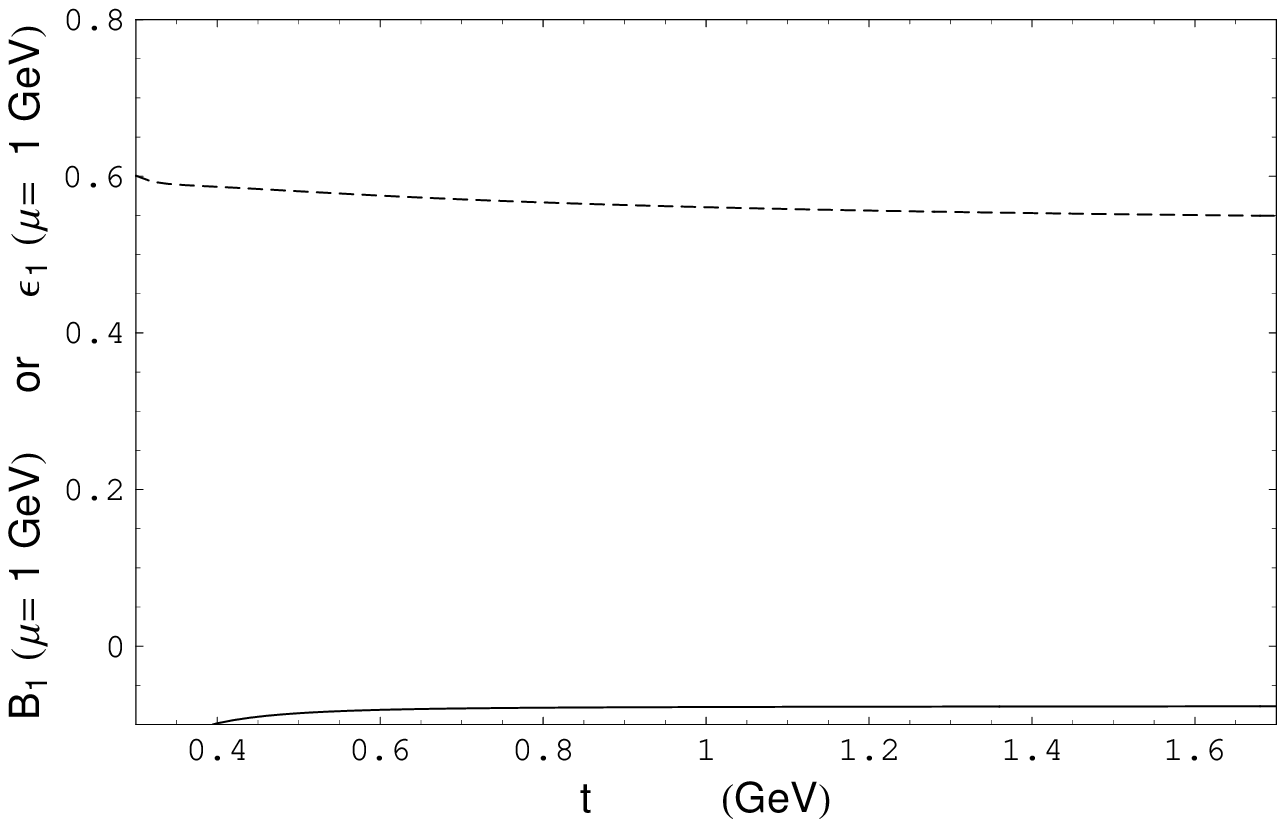,width=7.7cm,height=5.2cm}
    \ \epsfig{figure=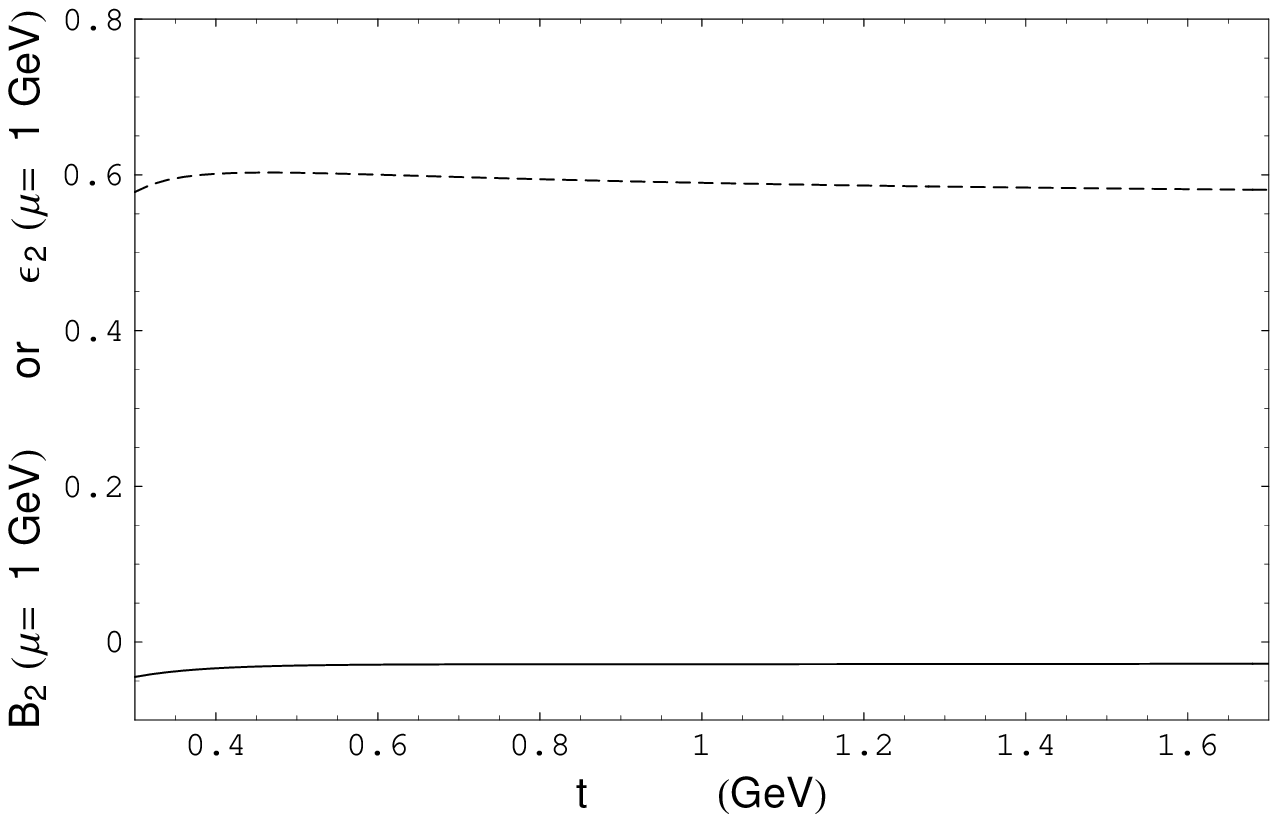,width=7.7cm,height=5.2cm}}
\vspace{0.2cm}
    \caption{{\small $B_i^v(\mu)$ and
$\epsilon_i^v(\mu)$ as a function $t$, where $B_i^v=8d_1^{v(i)}/9
+ 2d_2^{v(i)}/3,\ $ and $\epsilon_i^v=-4d_1^{v(i)}/27+
8d_2^{v(i)}/9$. The dashed and solid curves stand for $B_i^{v}$
and $\epsilon_i^{v}$, respectively. Here we have used
$\omega_0=1.2~ {\rm GeV}$ and Eq.~(\ref{omega3pt}).}}
\vspace{0.5cm}
\end{figure}

Substituting the above results into Eq.~(\ref{lowpara}) yields
\beq   \label{biei} B_1(m_b)=0.96\pm 0.04 + { O}(1/m_b)\,, &&
\qquad B_2(m_b)=0.95\pm 0.02 + { O}(1/m_b)\,, \nonumber\\
\epsilon_1(m_b)=-0.14\pm 0.01 +{ O}(1/m_b)\,,  && \qquad
\epsilon_2(m_b)=-0.08\pm 0.01 +{ O}(1/m_b)\,. \eeq It follows from
Eq.~(\ref{ratios}) that \beq && \frac{\tau (B^-)}{\tau(B_d)} =
1.11 \pm 0.02\,,   \nonumber \\ && \frac{\tau
(B_s)}{\tau(B_d)}\approx 1\,,  \nonumber\\ && \frac{\tau
(\Lambda_b)}{\tau (B_d)} = 0.99 - \Big(\frac{f_B}{185~{\rm
MeV}}\Bigr)^2 (0.007+0.020\, \tilde B)\, r \,, \eeq to the order
of $1/m_b^3$. Note that we have neglected the corrections of SU(3)
symmetry breaking to the nonspectator effects in $\tau
(B_s)/\tau(B_d)$. We see that the prediction for $\tau(B^-)/\tau
(B_d)$ is in agreement with the current world average:
$\tau(B^-)/\tau(B_d)$~=1.07$\pm$ 0.03 \cite{LEP}, whereas the
heavy-quark-expansion-based result for $\tau(B_s)/\tau(B_d)$
deviates somewhat from the central value of the world average:
$0.94\pm 0.04$. Thus it is urgent to carry out more precise
measurements of the $B_s$ lifetime. Using the existing sum rule
estimate for the parameter $r$ \cite{Col} together with $\tilde
B=1$ gives $\tau(\Lambda_b)/\tau(B_d)\geq 0.98$. Therefore, the
$1/m_b^3$ nonspectator corrections are not responsible for the
observed lifetime difference between the $\Lambda_b$ and $B_d$.

\section{Conclusions}
The nonspectator effects can be parametrized in terms of the
hadronic parameters $B_1$, $B_2$, $\epsilon_1$ and
$\epsilon_2$~\cite{recent}, where $B_1$ and $B_2$ characterize the
matrix elements of color singlet-singlet four-quark operators and
$\epsilon_1$ and $\epsilon_2$ the matrix elements of color
octet-octet operators. In OPE language, the prediction of $B$
meson lifetime ratios depends on the nonspectator effects of order
$16\pi^2/m_b^3$ in the heavy quark expansion. Obviously, the
shorter lifetime of the $\Lambda_b$ relative to that of the $B_d$
meson and/or the lifetime ratio $\tau(B_s)/\tau(B_d)$ cannot be
explained by the theory so far. It is very likely that local
quark-hadron duality is violated in nonleptonic decays.

  As emphasized in \cite{Cheng}, one should not be contented with the
agreement between theory and experiment for the lifetime ratio
$\tau(B^-)/ \tau(B_d)$. In order to test the OPE approach for
inclusive nonleptonic decay, it is even more important to
calculate the absolute decay widths of the $B$ mesons and compare
them with the data. From (\ref{nl}), (\ref{sl}), (\ref{biei}) and
considering the contributions of the nonspectator effects, we
obtain \beq \label{width} \Gamma_{\rm tot}(B_d) &=&
\,(3.61^{+1.04}_{-0.84}) \times 10^{-13}\,{\rm GeV}, \nonumber\\
\Gamma_{\rm tot}(B^-) &=& \,(3.34^{+1.04}_{-0.84}) \times
10^{-13}\,{\rm GeV}, \eeq noting that the next-to-leading QCD
radiative correction to the inclusive decay width has been
included. The absolute decay widths strongly depend on the value
of the $b$ quark mass. The problem with the absolute
decay width $\Gamma(B)$ is intimately related to the $B$ meson
semileptonic branching ratio ${\cal B}_{\rm SL}$. Unlike the
semileptonic decays, the heavy quark expansion in inclusive
nonleptonic decay is {\it a priori} not justified due to the
absence of an analytic continuation into the complex plane and
hence local duality has to be invoked in order to apply the OPE
directly in the physical region.

To conclude, we have derived in heavy quark effective theory the
renormalization-group improved sum rules for the hadronic
parameters $B_1$, $B_2$, $\epsilon_1$, and $\epsilon_2$ appearing
in the matrix element of four-quark operators. The results are
$B_1(m_b)=0.96\pm 0.04$, $B_2(m_b)=0.95\pm 0.02$,
$\epsilon_1(m_b)=-0.14\pm 0.01$ and $\epsilon_2(m_b)=-0.08\pm
0.01$ to the zeroth order in $1/m_b$. The resultant $B$-meson
lifetime ratios are $\tau(B^-)/\tau(B_d)=1.11\pm 0.02$ and
$\tau(B_s)/\tau(B_d)\approx 1$.

\acknowledgments The author thanks V. L. Chernyak for helpful
discussions and comments.  This work was supported in part by the
National Science Council of R.O.C. under Grant No.
NSC87-2112-M-001-048.
\thebibliography{99}
\def\bi{\bibitem}

\bibitem {BIGI}
I.I. Bigi, N.G. Uraltsev, and A.I. Vainshtein, Phys.\ Lett.\ B
{\bf 293}, 430 (1992) [{\bf 297}, 477(E) (1993)]; I.I. Bigi, M.A.
Shifman, N.G. Uraltsev, and A.I. Vainshtein, Phys.~
Rev.~Lett.~{\bf 71}, 496 (1993); A. Manohar and M.B. Wise, Phys.
Rev. {\bf D49}, 1310 (1994); B. Blok, L. Koyrakh, M. Shifman, and
A. Vainshtein, {\sl ibid.}, 3356 (1994); B. Blok and M. Shifman,
Nucl. Phys. {\bf B399}, 441 (1993); {\sl ibid.}, {\bf B399}, 459
(1993).

\bibitem{LEP} For updated world averages of $B$ hadron lifetimes,
see J. Alcaraz {\it et al.} (LEP $B$ Lifetime Group),
http://wwwcn.cern.ch/\~\,claires/lepblife.html.

\bi{Uraltsev} N.G. Uraltsev, \pl {\bf B376}, 303 (1996); J.L.
Rosner, \pl {\bf B379}, 267 (1996).

\bibitem{NS} M.~Neubert and C.T.~Sachrajda, Nucl.~Phys.~{\bf B483},
339 (1997); M.~Neubert, CERN-TH/97-148 [hep-ph/9707217].

\bibitem{Cheng} H.Y.~Cheng, Phys.~Rev.~D~{\bf 56}, 2783 (1997).

\bibitem{BLLS} M. S. Baek, J. Lee, C. Liu, and H.S. Song, Phys. Rev.
{\bf D57}, 4091 (1998).

\bi{Hokim} Q. Hokim and X.Y. Pham, \pl {\bf B122}, 297 (1983).

\bi{Bagan} E. Bagan, P. Ball, V.M. Braun, and P. Gosdzinsky, \pl {\bf B342},
362 (1995); {\sl ibid.} {\bf B374}, 363(E) (1996); E. Bagan, P. Ball, B.
Fiol, and P. Gosdzinsky, {\sl ibid.} {\bf B351}, 546 (1995); M. Lu, M. Luke,
M.J. Savage, and B.H. Smith, Phys. Rev. {\bf D55}, 2827 (1997).

\bi{CT} H.Y. Cheng and B. Tseng, hep-ph/9803457.

\bibitem{Keum} Y.Y. Keum and U. Nierste, Phys.\ Rev.\ {\bf D57}, 4282 (1998).

\bibitem {SV1}
M.A. Shifman and M.B. Voloshin, Sov.\ J.\ Nucl.\ Phys.\ {\bf 41},
120 (1985).

\bi{HY} H.-Y. Cheng and K.-C. Yang, IP-ASTP-03-98
[hep-ph/9805222], to appear in Phys. Rev. D.

\bibitem{yang1} K.-C. Yang, Phys.~Rev.~D~{\bf 57}, 2983 (1998);
K.-C. Yang, W-Y. P. Hwang, E.M. Henley, and L.S. Kisslinger, {\sl
ibid.}~D~{\bf 47}, 3001 (1993); K.-C. Yang and W-Y. P. Hwang, {\sl
ibid.}~D~{\bf 49}, 460 (1994).

\bibitem{Neubert2} M. Neubert, Phys.~Rev.~D {\bf 45}, 2451 (1992);
{\bf 46}, 1076 (1992).

\bibitem{BB} P. Ball and V.M. Braun, Phys.~Rev.~D~{\bf 49}, 2472 (1994).

\bibitem {Chern}
V. Chernyak, Nucl. Phys. {\bf B457}, 96 (1995).

\bibitem{lattice} C. Bernard {\it et al.,} hep-ph/9709328.

\bibitem{Ball} P. Ball, Nucl.~Phys.~{\bf B421}, 593 (1994).

\bi{Abe} CDF Collaboration, F. Abe {\it et al.,} Phys. Rev. {\bf D57}, 5382
(1998).

\bi{Col} P. Colangelo and F. De Fazio, \pl {\bf B387}, 371 (1996).

\bi{recent} For recent works, see M. Di Pierro and C.T. Sachrajda,
Nucl. Phys. B534, 373, (1998), and D. Pirjol and N. Uraltsev,
UND-HEP-98-BIG-03 [hep-ph/9805488].

\end{document}